\documentstyle[12pt]{article}
\def\<{{\langle}}
\def\>{{\rangle}}

\def\be{\begin{equation}}
\def\ee{\end{equation}}
\def\ba{\begin{eqnarray}}
\def\ea{\end{eqnarray}}
\def\mref#1{Eq.(\ref{Eq:#1})}

\def\mlab#1{\label{Eq:#1}}
\def\mlabf#1{\label{Eq:#1}}

\def\half{\frac{1}{2}}
\def\to{\rightarrow}
\def\nn{\nonumber\\}

\def\PR#1#2#3 {{\it Phys. Rev. }{\bf D#1} #2 {(#3)} }
\def\PRL#1#2#3 {{\it Phys. Rev. Lett. }{\bf #1} #2 {(#3)} }
\def\PL#1#2#3 {{\it Phys. Lett. }{\bf #1} #2 {(#3)}  }
\def\AP#1#2#3 {{\it Ann, Phys. }{\bf #1} #2 {(#3)} }
\def\ZP#1#2#3 {{\it Z. Phys. }{\bf #1} #2 {(#3)} }
\def\NP#1#2#3 {{\it Nucl. Phys. }{\bf #1} #2 {(#3)}  }
\def\MPL#1#2#3 {{\it Mod. Phys. Lett.}{\bf #1} #2 {(#3)}  }
\def\NC#1#2#3 {{\it Nuov. Cimm. }{\bf #1} #2 {(#3)}  }
\def\PREP#1#2#3 {{\it Phys. Report }{\bf #1} #2 {(#3)}  }
\def\PROG#1#2#3 {{\it Prog. Theor. Phys. }{\bf #1} #2 {(#3)}   }

\def\cp{{\bf CP}}

\def\sq2{{1\over{\sqrt{2}}}}

\def\g5{\gamma_5}

\renewcommand{\Im}{\mbox{Im}\,}

\begin{document}
\input epsf.sty
\renewcommand{\thefootnote}{\fnsymbol{footnote}}


\begin{flushright}
UND-HEP-99-BIG\hspace*{.2em}06\\
DPNU-99-30\\
hep-ph/9909479\\
\end{flushright}
\vspace{.3cm}
\begin{center} \Large 
{\bf On the Other Five Unitarity Triangles}
\end{center}
\vspace*{.3cm}
\begin{center} {\Large 
I. I. Bigi $^{a}$, A. I. Sanda $^{b}$}\\
\vspace{.4cm}
{\normalsize 
$^a${\it Physics Dept.,
Univ. of Notre Dame du
Lac, Notre Dame, IN 46556, U.S.A.}\\
$^b$ {\it  Physics Dept., Nagoya University, 
Nagoya 464-01,Japan}}
\\
\vspace{.3cm}
e-mail addresses:\\
{\it bigi@undhep.hep.nd.edu, 
sanda@eken.phys.nagoya-u.ac.jp} 
\vspace*{.4cm}

{\Large{\bf Abstract}}\\
\end{center} 
A comprehensive program of \cp~studies in heavy flavour 
decays has to go beyond observing large \cp~asymmetries in 
nonleptonic $B$ decays and finding that the sum of the three 
angles of the unitarity triangle is consistent with 180$^{\circ}$. 
There are many more correlations between observables 
encoded in the KM matrix; those can be expressed through five 
unitarity 
triangles in addition to the 
one usually considered. To test the completeness of the KM 
description one has to obtain a highly overconstrained data set 
sensitive to ${\cal O}(\lambda ^2)$ effects with 
$\lambda = \sin \theta _C$. Those fall into two categories: 
(i) Certain large angles agree to leading order only, yet 
differ in order $\lambda ^2$ in a characteristic way. 
(ii) Two observable angles are -- for reasons specific to the 
KM ansatz -- ${\cal O}(\lambda ^2)$ 
and ${\cal O}(\lambda ^4)$ thus generating an asymmetry of 
a few percent and of about 0.1 \%, respectively. The former can be 
measured in $B_s \to \psi \eta , \, \psi \phi$ 
{\em without} hadronic uncertainty, the latter in 
Cabibbo suppressed $D$ decays. 
The intervention of New Physics could boost these 
effects by an order of magnitude. 
A special case is provided by $D^+ \to K_{S,L}\pi ^+$ 
vs. $D^- \to K_{S,L}\pi ^-$. 
Finally, \cp~asymmetries involving 
$D^0 - \bar D^0$ oscillations could reach observable levels 
only due to New Physics.  

\vspace*{.2cm}
\vfill
\noindent
\vskip 5mm

\tableofcontents 
\section{Overview}
\label{INTRO}

It has become customary to talk about {\em the} KM unitarity 
triangle as the one that represents the relation 
\be 
V^*_{ub} V_{ud}  + V^*_{cb} V_{cd}  + V^*_{tb} V_{td}  = 
\delta _{bd} = 0
\mlab{TRI5}
\ee
This triangle plays a central role in $B$ decays with the three 
terms in \mref{TRI5} controlling $b\to u$, $b\to c$ transitions 
and  
$B_d - \bar B_d$ oscillations, respectively. It also has the 
important feature 
that its sides 
have comparable lengths, namely of order $\lambda ^3$, where 
$\lambda = {\rm sin} \theta _C$. Its three angles 
$\phi _{1,2,3}$  
\be 
\phi _1 = \pi - 
{\rm arg} \left( \frac{- V^*_{tb} V_{td}}{- V^*_{cb} V_{cd}} 
\right) \; , \; 
\phi _2 = 
{\rm arg} \left(\frac{ V^*_{tb} V_{td}}{- V^*_{ub} V_{ud}}
\right)  \; , \; 
\phi _3 =  {\rm arg} 
\left( \frac{ V^*_{ub} V_{ud}}{- V^*_{cb} V_{cd}} \right) 
\; ,  
\mlab{ANGLES1} 
\ee
are therefore 
naturally large -- as are the \cp~asymmetries they 
generate in $B$ decays. Beyond this qualitative observation 
the KM ansatz unequivocally states that these \cp~asymmetries 
are such that 
\be 
\phi _1 + \phi _2 + \phi _3 = 180 ^{\circ} 
\mlab{180}
\ee
holds. We anticipate that in the first round of measurements 
\cp~violation will be established in the beauty sector, 
presumably in $B_d \to \psi K_S$ 
\cite{OPALCDF}; 
in a second round all three 
angles will be extracted with some degree of accuracy and 
\mref{180} will be checked empirically. Yet -- and this is the 
main message of this paper -- a {\em complete} program has 
to move well beyond this stage both quantitatively and 
qualitatively:   
\begin{itemize}
\item 
A dedicated effort has to be undertaken to determine the 
values of $\phi _{1,2,3}$ as accurately as possible: 
\begin{itemize}
\item 
The only practical limitation on determining $\phi _1$ is of an 
experimental nature, and an accuracy of better than 5\% appears 
attainable. 
\item 
It should be possible in the long run to determine 
$|V_{ub}/V_{cb}|$ and $|V_{td}/V_{cb}|$ with 5\% and 10\% 
accuracy, respectively. It would allow us to construct the 
triangle of \mref{TRI5} with less than 10\% uncertainty. Comparing 
the values {\em measured} for these angles with those 
{\em inferred} from the triangle then provides a highly sensitive 
probe of New Physics. 
\item 
Extracting $\phi_2$ and $\phi_3$ with similar precision will pose 
quite a challenge for theoretical and other reasons extensively 
discussed in the literature \cite{KHOZE,BABAR,BOOK}. Whether 
ultimately the 5\% accuracy 
level can be reached here is far from certain at the moment; 
we want to emphasize that it is a highly desirable goal deserving a 
dedicated effort. 
\item 
In that context we would like to sound the following note.  The 
effective branching ratios for the interesting $B$ decays are 
small basically since so many channels are available. Once the 
nontrivial investment has been made to accumulate sufficient 
statistics in these modes one can turn this vice into a virtue: 
one can learn valuable lessons about the hadronization process 
by analyzing the multitude of channels driven by a mere 
handful of quark level transitions. We are optimistic that such 
studies will improve our theoretical understanding very 
significantly by trial and error.  
\end{itemize}
\item 
To obtain the desired accuracy one wants to combine the 
measurements on certain decay modes to enhance the 
statistics; inferring consistent values from different 
channels would also demonstrate that theoretical control 
has indeed been established. 
\item 
Yet a closer look at the weak parameters controlling 
the asymmetries in these modes reveal that they agree to leading 
order in $\lambda$ only; in higher orders they differ in a 
way that is very specific to the KM ansatz; a deviation reveals 
the intervention of New Physics. 
\item 
In some cases it is important to compare angles extracted from  
different asymmetries even if within the KM ansatz 
they have to coincide for all practical purposes. This 
represents a meaningful probe for New Physics in particular 
if one channel is dominated by a tree amplitude and the other 
by a Penguin amplitude \cite{SONI}.   
\end{itemize} 

A comprehensive program 
sensitive to such higher order effects 
has to be undertaken to probe the completeness of the KM 
description and to exploit the discovery potential 
to the fullest; it will have to {\em include charm studies}. 
This can be best discussed in terms of the 
other five triangle relations that follow from the 
$3\times 3$ KM matrix being unitary 
\cite{XING}. 

While all six triangles possess the same 
area, their shapes are quite different.   They can be 
grouped into three categories of two triangles each: 
\begin{enumerate}
\item 
In addition to the {\em bd} triangle of \mref{TRI5} there is 
another one where the lengths of 
all sides are of order $\lambda ^3$: it is represented 
by the relation 
\be 
V^*_{ud} V_{td}  + V^*_{us} V_{ts} + V^*_{ub} V_{tb} =
\delta _{tu} = 0 
\; \; \; \; \;  \hat = \; \; \; \; \; tu \; \; {\rm triangle} 
\mlab{TRI6}
\ee
To leading order in $\lambda$ -- 
where $V_{ts} = - V_{cb}$ holds -- it coincides with the first one 
given by \mref{TRI5}. 
\item 
Two triangles have a squashed appearance with 
the lengths of two sides of order $\lambda ^2$ and the third 
one of order $\lambda ^4$: 
\be 
V^*_{us} V_{ub}  + V^*_{cs} V_{cb}  + V^*_{ts} V_{tb} = 
\delta _{bs} = 0 
\; \; \; \; \;  \hat = \; \; \; \; \; bs \; \; {\rm triangle} 
\mlab{TRI3}
\ee 
\be 
V^*_{td} V_{cd} + V^*_{ts} V_{cs} + V^*_{tb} V_{cb} =
\delta _{tc} = 0 
\; \; \; \; \;  \hat = \; \; \; \; \; tc \; \; {\rm triangle} 
\mlab{TRI4}
\ee
\item 
The remaining two triangles are even more extreme with 
the lengths of two sides of order $\lambda$ and the third 
one of order $\lambda ^5$:
\be 
V^*_{ud} V_{us} + V^*_{cd} V_{cs} + V^*_{td} V_{ts} = 
\delta _{sd} = 0 
\; \; \; \; \;  \hat = \; \; \; \; \; sd \; \; {\rm triangle} 
\mlab{TRI1}
\ee 
\be 
V^*_{ud} V_{cd} + V^*_{us} V_{cs} + V^*_{ub} V_{cb} = 
\delta _{cu} = 0 
\; \; \; \; \;  \hat = \; \; \; \; \;  cu \; \; {\rm triangle} 
\mlab{TRI2}
\ee

\end{enumerate} 
Obviously we want to determine the fundamental KM 
parameters as precisely as possible in the hope that a 
future more comprehensive theory will explain them. 
As is well-known the KM matrix contains four 
{\em independant} quantities in terms of which a host of 
observables is described. Thus there exist numerous correlations 
between these observables as expressed through the 
geometry of the six triangles. For example -- as described 
below -- to leading order 
in $\lambda$ one angle in four of the triangles coincides, yet 
differs in higher orders in a characteristic way. These higher 
order effects have to be studied when testing the completeness 
of the KM description thoroughly and sensitively  
as a probe for the presence of New Physics and the salient 
features of the latter. This will be achieved by measuring 
as many sides and angles of the six KM triangles as possible 
and as precisely as possible to obtain a highly overconstrained 
data set. 

In this note we will 
\begin{itemize}
\item 
show how the angles of the {\em other} triangles can be 
interpreted, 
\item 
discuss how several of them can be measured which will 
\item 
yield a data set with highly overconstrained information 
on the KM parameters 
\item 
that probes the completeness of the KM description thoroughly. 
\end{itemize}
Our message consists of two main parts:
\begin{enumerate}
\item 
The goal should be to determine as many angles as accurately 
as possible. The benchmark here is the 5\% accuracy level. 
\item 
A full program on heavy flavour decays has to include 
dedicated searches for \cp~asymmetries in charm decays. 
A 0.1\% sensitivity here would describe a complete program. 
\end{enumerate} 
While it is not clear at present whether these benchmarks 
can be reached, they do not appear to be 
unrealistic. 

The paper will be organized as follows: in Sect. 
\ref{INTERPRETATION} we describe how the angles of the 
KM triangles are in principle related to observables; in 
Sect. \ref{BS} and \ref{CHARM} we discuss in some detail 
two of those triangles, namely the $bs$ triangle with 
$B_s \to \psi \eta$ as its central transition and the 
$cu$ triangle describing charm decays; in 
Sect. \ref{GATES} we list gateways through which New Physics 
could enter before summarizing 
in Sect. \ref{SUMMARY}.  

\section{Interpretation of the Six Triangles 
\label{INTERPRETATION}
}

\subsection{Terminology and Notation}

For numbering the angles of the various triangles we follow 
the same systematics as for the first triangle, \mref{TRI5}: 
in triangle $mn$ we call $\phi_i^{mn}$ the angle {\em opposite} 
the side from the $i$th family, i.e. ${\bf V}_{mi} {\bf V}_{ni}^*$ 
or ${\bf V}_{im}{\bf V}_{in}^*$ . 
Then we rescale the triangles such that their base has unit length.

{\em Direct} \cp~asymmetries are described by $\Delta B=1$ 
or $\Delta C=1$ amplitudes alone. \cp~violation in 
$B^0 - \bar B^0$ oscillations as probed in wrong-sign 
semileptonic $B$ decays is controled by a $\Delta B=2$ 
amplitude. \cp~asymmetries involving oscillations arise 
from the interplay of $\Delta B= 1 \& 2$ or $\Delta C= 1 
\& 2$ transitions. As long as one studies such an 
asymmetry in a {\em single} channel only, its assignment 
to the $\Delta B = 1$ or $\Delta B =2$ sector - 
in the first [second] case it would be referred to as 
direct 
[indirect] \cp~violation -- is 
arbitrary, since a phase rotation 
in the quark fields would shift it from one to the other 
sector. Only if two $B_d$ (or $B_s$) channels exhibit 
\cp~asymmetries that differ beyond their \cp~parities, 
has direct \cp~violation been established unequivocally. 

Four classes of $\Delta B=1$ or $\Delta C=1$ amplitudes 
have to be distinguished with respect to the theoretical 
control one has over them and their potential to 
be affected by New Physics: decays that are   
\begin{itemize} 
\item 
dominated by a single tree level process,  
\item 
receiving contributions by two tree level 
reactions,  
\item 
being affected significantly by a Penguin process in addition 
to the tree level one, or 
\item 
dominated by Penguin reactions.

\end{itemize}

To make the unitarity constraints for 
three families explicit we will also employ the Wolfenstein 
expansion 
of the KM matrix up to order $\lambda ^4$ [$\lambda ^6$] for the 
real [imaginary] parts:  
\be 
{\bf V}_{CKM} = 
\left( 
\begin{array}{ccc} 
1 - \frac{1}{2} \lambda ^2 & \lambda & 
A \lambda ^3 (\rho - i \eta + \frac{i}{2} \eta \lambda ^2) \\
- \lambda & 1 - \frac{1}{2} \lambda ^2 - i \eta A^2 \lambda ^4 & 
A\lambda ^2 (1 + i\eta \lambda ^2 ) \\ 
A \lambda ^3 (1 - \rho - i \eta ) 
& - A\lambda ^2 & 1 
\end{array}
\right) 
+ {\cal O}(\lambda ^6) 
\mlab{WOLFKM}
\ee 
where 
\be 
\lambda \equiv {\rm sin} \theta _C
\ee

\subsection{On Determining Angles}
\begin{enumerate}
\item 
The rescaled $bd$ triangle is expressed through 
\be 
1 + \frac{ V^*_{tb} V_{td}}{ V^*_{cb} V_{cd}} + 
\frac{ V^*_{ub} V_{ud}}{ V^*_{cb} V_{cd}} = 0 
\mlab{TRI1A}
\ee 
As is well known, its angles $\phi _1$, $\phi _2$ and $\phi _3$  
can be measured in $B$ decays: 
\begin{itemize}
\item 
$\phi _1$ can be determined through the time dependant \cp~asymmetry 
in $B_d (t) \to \psi \pi \pi$, $D^{(*)} \bar D^{(*)}$ controlled by 
\be 
{\rm Im}\left( \frac{q}{p} \overline \rho _f \right) \equiv 
{\rm Im} \left( \frac{q}{p} \frac{T(\overline B_d \to f)}{T(B_d \to 
f)} \right) \; . 
\ee 
Up to tiny corrections of order 
$\lambda ^4$ this can be achieved much more easily 
in $B_d \to \psi K_S$ 
\cite{BS1} where we find: 
$$  
{\rm Im} \left( \frac{q}{p} \overline \rho _{\psi K_S}\right)  = 
{\rm Im}\left( \frac{V^*_{td} V_{tb}}{V^*_{tb}V_{td}}
\frac{V^*_{cb} V_{cs}}{V^*_{cs}V_{cb}}\right)  = 
\sin 2\phi _1   
$$  
\be 
\simeq  
\frac{2\eta (1 - \rho )}{(1 - \rho )^2 + \eta ^2} 
\left[  1 - \lambda ^2 \left( 1 - \rho - 
\frac{\eta ^2}{1- \rho } \right) 
\right]  + {\cal O}(\lambda ^4) 
\mlab{PSIKS}
\ee 
$\phi _1$ could be inferred also from the difference 
in $B_d(t) \to D_{+[-]} \pi ^+ \pi ^-$ vs.  
$\bar B_d (t) \to D_{+[-]} \pi ^+ \pi ^-$ where $D_{+[-]}$ denotes 
the 
\cp~even [odd] combination of $D^0$ and $\bar D^0$ 
\cite{CARTER}  
\footnote{The discerning reader will have noticed that 
$\frac{V^*_{td} V_{tb}}{V^*_{tb}V_{td}}
\frac{V^*_{cb} V_{ud}}{V^*_{ud}V_{cb}} $ is {\em not} 
invariant under phase rotations of the $c$ and $u$ fields. 
That does not pose a problem here since one analyses the 
transitions in terms of the \cp~eigenstates $D_{\pm}$ 
rather than the flavour eigenstates $D^0$ and $\bar D^0$. 
An analogous situation holds arises for 
$B_d \to \psi K_S$, see \mref{PSIKS}.}.  
\ba 
D_+ &\to& K^+ K^-, \; \pi ^+ \pi ^-, ... \nn 
D_- &\to& K_S \pi ^0, \; K_S \eta , ...  \; . 
\ea 
since 
\be 
\sin 2\phi _1 = {\rm Im} 
\frac{V^*_{td} V_{tb}}{V^*_{tb}V_{td}}
\frac{V^*_{cb} V_{ud}}{V^*_{ud}V_{cb}} 
+ {\cal O}(\lambda ^4)   
\ee 
holds. 
However it appears unlikely that such an analysis could yield 
a competitive extraction, and in any case we view it as 
quite unlikely that New Physics could intervene to induce 
a significant difference between the two values. 

Here we can also illustrate our previous comment on whether 
an observed \cp~violation originated from $\Delta B=2$ 
or $\Delta B=1$ dynamics: the Wolfenstein 
parametrization, \mref{WOLFKM}, would suggest that the 
asymmetry in $B_d \to \psi K_S$ is due to 
arg $V_{td} \neq 0$ which affects the $B_d - \bar B_d$ 
oscillations; yet a phase rotation 
$(t,b) \to (t,b) e^{-i\phi _1}$ would make [keep] 
$V_{td} [V_{tb}]$ real with the large complex phase 
re-surfacing in $V_{cb}$ which appears in the 
$\Delta B=1$ amplitude.
\item 
There is the intriguing possibility to extract 
$\phi _1$ from the asymmetry in 
$\bar B_d (t) \to \phi K_S$ vs. $B_d (t) \to \phi K_S$ 
\cite{SONI}. The $\Delta B=1$ transition 
amplitude is generated mainly 
by Penguin operators customarily expressed by 
\cite{PDG2000} 
\be 
T(\bar B_d \to \phi K_S) = 
V_{cb}V_{cs}^* \left( P_c - P_t \right) + 
V_{ub} V_{us}^* \left( P_u - P_t \right) \; ; 
\mlab{PENGUIN} 
\ee 
$P_q$ contains the Penguin operator with an internal quark 
$q$. It follows from the GIM mechanism as implemented 
through a unitary KM matrix that only the differences 
between the operators with different internal quarks 
enter. The contribution from $P_c - P_t$ is leading with the 
one from $P_u - P_t$ representing an ${\cal O}(\lambda ^2)$ 
correction. 

The asymmetry is then controlled by 
\be 
{\rm Im}\frac{V_{td}^*V_{tb}}{ V_{tb}^*V_{td}}
\frac{V_{cb}^*V_{cs}}{ V_{cs}^*V_{cb}} = 
\sin 2\phi _1 \; . 
\ee 
The important point to remember here is 
that a Penguin operator constituting a quantum correction 
is sensitive to New Physics operating at high mass scales. 
Thus it is quite conceivable that the asymmetries in 
$\bar B_d (t) \to \phi K_S$ 
and $\bar B_d \to \psi K_S$ would yield very 
different values for $\phi _1$.  
A limiting factor in the theoretical interpretation of 
such a probe is provided by the $P_u - P_t$ 
contribution and the degree of computational control 
that can be established over it.   

\item 
$\phi _2$ can be extracted from $B_d(t) \to \pi 's$: 
$$  
\sin 2\phi _2 = {\rm Im} 
\frac{V^*_{td} V_{tb}}{V^*_{tb}V_{td}}
\frac{V^*_{ub} V_{ud}}{V^*_{ud}V_{ub}} 
\simeq \frac{2\eta}{[(1-\rho )^2 +\eta ^2]
[\rho ^2 + \eta ^2 ( 1 - \lambda ^2)]} \cdot 
$$ 
\be 
\left[ \rho - \rho ^2 -\eta ^2 + 
\lambda ^2\left( \rho ^2 - \frac{1}{2} \rho ^3 - \frac{1}{2}\rho 
+ \frac{1}{2} \eta ^2 (1 - \rho ) \right) \right] 
+ {\cal O}(\lambda ^4)   
\ee  
While the intervention of Penguin transitions poses a serious 
problem here, we are optimistic 
that the various methods suggested to unfold 
the Penguin pollution will succeed at least in the long run 
\cite{BABAR}. 
Yet whether an accuracy of $\lambda ^2 \simeq 5\%$ or even 
better can be achieved is quite unclear at present. 
\item 
$\phi _3$ induces a direct \cp~asymmetry in 
$B^+ \to D^0/\bar D^0 K^+$ vs. 
$B^- \to D^0/\bar D^0 K^-$ \cite{BSPHI3,BS4}; 
$D^0/\bar D^0$ denotes a {\em coherent} superposition 
of $D^0$ and $\bar D^0$. This situation arises when 
the neutral charm meson decays into a \cp~eigenstate: 
$D^0/\bar D^0 = D_{\pm}$. Measuring also 
$B^{\pm} \to D^0 K^{\pm}$, $\bar D^0 K^{\pm}$ -- i.e., 
when the neutral charm meson decays in a flavour 
specific manner -- allows us the 
determine the relevant hadronic quantities and thus 
extract $\phi _3$ {\em cleanly} \cite{GRONAU}. 
The asymmetry depends on  
\ba 
{\rm Im} \left( \frac{V^*_{cs}V_{ub}}{V^*_{us}V_{cb}} \right) 
&=&  
 {\rm Im} \left( \frac{V^*_{ud}V_{ub}}{-V^*_{cd}V_{cb}} \right) 
+ {\cal O} (\lambda ^4)  = \sin \phi _3   
\nn 
&=& - \eta 
\left[ 1 - \lambda ^2 \left( \frac{1}{2} - \rho \right) \right] 
+ {\cal O}(\lambda ^4) 
\ea
A probably more practical variant of this method is 
to rely on a final state that is common to $D^0$ and 
$\bar D^0$ decays due to the presence of a doubly 
Cabibbo suppressed transition: 
$D^0 \rightarrow K^+ \pi ^-, \, 
K^- \pi ^+ \leftarrow \bar D^0$ 
\cite{ATWOOD}.  
\item 
An `oblique' angle can be studied by comparing 
\cite{KHOZE,BS4} 
\be 
\overline B_d (t) \to D^{0 (*)} K_S \; \; \; vs. \; \; \; 
B_d (t) \to \overline D^{0 (*)} K_S \; ; 
\mlab{OBLIQUE1} 
\ee 
their difference is controlled by 
Im $\frac{V_{tb} V^*_{td}}{V^*_{tb} V_{td}} 
\frac{V^*_{ub} V_{cs}}{V_{cb}V^*_{us}} $ with 
\be 
{\rm Im} \frac{V_{tb} V^*_{td}}{V^*_{tb} V_{td}} 
\frac{V^*_{ub} V_{cs}}{V_{cb}V^*_{us}} \simeq 
\left|  \frac{V_{ub} V_{cs}}{V_{cb}V_{us}} \right| 
\sin \left( {\rm arg} \frac{V^*_{td}}{V_{cb}} 
\frac{V^*_{ub}}{V_{td}}  
\right) \simeq \left|  \frac{V_{ub} V_{cs}}{V_{cb}V_{us}} \right|  
\sin (\phi _1 - \phi _2) 
\ee
Analogously the difference in 
\be 
\overline B_d (t) \to \overline D^{0 (*)} K_S \; \; \; vs. \; \; \; 
B_d (t) \to  D^{0 (*)} K_S \; ; 
\mlab{OBLIQUE2} 
\ee 
 is controlled by 
\be 
{\rm Im} \frac{V_{tb} V^*_{td}}{V^*_{tb} V_{td}} 
\frac{V^*_{cb} V_{us}}{V_{ub}V^*_{cs}} \simeq 
\left|  \frac{V_{us} V_{cb}}{V_{ub}V_{cs}} \right| 
\sin \left( {\rm arg} \frac{V^*_{td}}{V^*_{cb}} 
\frac{V_{ub}}{V_{td}}  
\right) \simeq \left|  \frac{V_{us} V_{cb}}{V_{ub}V_{cs}} \right|  
\sin (\phi _1 - \phi _2) 
\ee
Comparing the two asymmetries in \mref{OBLIQUE1} and 
\mref{OBLIQUE2} allows to infer sin$(\phi _1 - \phi _2)$ 
cleanly. 
\end{itemize} 
That (at least) two angles can be observed in 
$B_d$ transitions is not surprising since \mref{TRI1A} 
represents the $bd$ element of 
$V^{\dagger}_{KM} V_{KM}$. We will {\em not} add a 
superscript $bd$ to these angles of the {\em standard} 
unitarity triangle. 

Obviously only two of the three angles $\phi _1$, $\phi _2$ 
and $\phi _3$ are independent of each other. In the 
Wolfenstein representation, \mref{WOLFKM}, only $V_{ub}$ and $V_{cb}$ have 
large phases; then one would be tempted to argue that 
$\phi _1 \simeq - {\rm arg} V_{td}$ and 
$\phi _3 \simeq - {\rm arg} V_{ub}$ are elementary whereas 
$\phi _2 \simeq  + {\rm arg} V_{td} + {\rm arg} V_{ub}$ 
is composite. Yet that would be fallacious: 
changing the phase of the $b$ and $t$ quark fields 
by $\phi _3$ and $\phi _1$, respectively, makes 
$V_{ub}$ and $V_{td}$ basically real; 
the large 
phases now surface in $V_{cb}$, $V_{tb}$ and $V_{ts}$. 
The values of $\phi _1$, $\phi _2$, $\phi _3$ do not 
change, of course, since they are phase invariant, 
yet their make-up does  
\be 
\phi _1 \simeq {\rm arg} V_{tb} - {\rm arg} V_{cb}\; , \; \; 
\phi _2 \simeq \pi - {\rm arg} V_{tb} \; , \; \; 
\phi _3 \simeq {\rm arg} V_{cb}\; ,  
\ee  
which would make $\phi _{2,3}$ look elementary and 
$\phi _1$ composite - contrary to the appearance in the 
original parametrization! Thus the three angles have to 
be viewed on the same level - at least until one comes to 
understand their dynamical origin.   

Yet as long as this is kept in mind it is very convenient 
to express the angles 
through those KM parameters that 
exhibit a complex phase in the Wolfenstein expansion; 
this will be denoted by the superscript $W$: 
\be 
\phi_1 = - {\rm arg} \, V_{td}^W - {\rm arg} \, V_{cb}^W 
\; , \; 
\phi _2 = \pi  + {\rm arg} \, V_{td}^W + 
{\rm arg} \, V_{ub}^W 
\; , \; 
\phi _3 = - {\rm arg} \, V_{ub}^W + {\rm arg} \, V_{cb}^W 
\ee 
\item 
The rescaled $tu$ triangle 
\be 
1 + \frac{V_{td} V^*_{ud}}{V_{ts}V^*_{us}} + 
\frac{V_{tb} V^*_{ub}}{V_{ts}V^*_{us}} = 0 
\mlab{TRI1B}
\ee 
\be 
\phi _1^{tu} = 
 {\rm arg} \frac{ -V^*_{ub}V_{tb}}{ V^*_{us}V_{ts}}
\; , \; \; 
\phi _2^{tu} = 
 {\rm arg} \frac{-V^*_{ud}V_{td}}{ V^*_{ub}V_{tb}}\; , \; \; 
\phi _3^{tu} = 
\pi -{\rm arg} \frac{ V^*_{ud}V_{td}}{ V^*_{us}V_{ts}} 
\ee 
\be 
\phi _3^{tu} = - {\rm arg} \, V_{td}^W 
\; , \; 
\phi _2^{tu} = 
\pi + {\rm arg} \, V_{td}^W +{\rm arg} \, V_{ub}^W 
\; , \; 
\phi _1^{tu} = - {\rm arg} \, V_{ub}^W  
\ee
coincides with the first one through order $\lambda$: 
\be 
\phi _3^{tu} = \phi _1 + {\cal O}(\lambda ^2), \; 
\phi _2^{tu} = \phi _2 , \; 
\phi _1^{tu} = \phi _3 + {\cal O}(\lambda ^2) 
\ee
In order $\lambda ^2$ the shapes of the 
$bd$ and $tu$ triangles differ since 
$\frac{V_{cb} + V_{ts}}{|V_{cb}|} = i\eta \lambda ^2$. 
It would be instructive if one could measure the angles of the 
two triangles separately and check whether they agree to 
leading order  {\em and} their ${\cal O}(\lambda ^2)$ differences 
follow the pattern described by the KM matrix. This seems, 
however, to be quite unrealistic even beyond the question of 
statistical precision. For this triangle represents the $tu$ element 
of $V_{KM}^{\dagger} V_{KM}$. Its geometry could be probed 
directly in 
top {\em hadron} transitions; alas top quarks decay before 
they can hadronize 
\cite{TOP}
\footnote{Even if top hadrons like $T^0 = [t \bar u]$ existed, such 
prospects would presumably be quite academic due to slow 
$T^0 - \bar T^0$ oscillations etc.}. 

There is one noteworthy exception, though: 
The angle $\phi _3^{tu}$ can be 
extracted from the \cp~asymmetry in 
$B_s(t) \to K_S \rho ^0$. For the latter depends on 
\be 
{\rm Im}\frac{V^*_{ts}V_{tb}}{V^*_{tb}V_{ts}} 
\frac{V^*_{ub}V_{ud}}{V^*_{ud}V_{ub}} = 
\sin 2\phi _3^{tu} \simeq 
\frac{2\rho \eta \left( 1 - \half \lambda ^2 \right)}
{\rho ^2 + \eta ^2 (1 - \lambda ^2)} 
\ee 
\be 
{\rm tg} \phi _3^{tu} = \frac{\eta}{\rho} 
\left( 1 - \frac{\lambda ^2}{2} \right) \; . 
\mlab{TU3}
\ee 

In such an analysis one has to overcome the theoretical 
challenge of unfolding the (Cabibbo suppressed) Penguin 
contribution. Due to the anticipated tiny branching ratio 
this is unlikely to be a practical method for a precise 
measurement. 

\item 
The $bs$ triangle 
\be 
1+ \frac{V^*_{cs}V_{cb}}{V^*_{ts}V_{tb}} + 
\frac{V^*_{us}V_{ub}}{V^*_{ts}V_{tb}}  = 0 
\ee  
is qualitatively different as pointed out before: it is a squashed 
triangle with one angle -- $\phi _1^{bs}$ -- much smaller 
than the other two: 
\be 
\phi _1^{bs} = \pi + {\rm arg} 
\left( \frac{V^*_{cs}V_{cb}}{V^*_{ts}V_{tb}} \right) 
\simeq  \lambda ^2 \eta \simeq  0.05 \cdot \eta 
\ee
Since it is a novel angle, we denote it specially 
\cite{PDG2000} 
\be 
\phi _1^{bs} \equiv \chi 
\mlab{CHI}
\ee
As discussed in the next section $\chi$ can be determined in 
$B_s \to \psi \eta$, $\psi \phi$ 
\cite{BS1,BS2,SILVA}. 

With $\phi _1^{bs}$ being so small, the other two angles 
are practically complementary:
\be 
\phi _3^{bs} \equiv 
{\rm arg} 
\left( \frac{ V^*_{us}V_{ub}}{- V^*_{cs}V_{cb}}\right)
\simeq \pi - \phi _2^{bs} \; \; , \; \; 
\phi _2^{bs} = 2\pi -{\rm arg} 
\left( \frac{- V^*_{us}V_{ub}}{ V^*_{ts}V_{tb}} \right)
\ee
or in our usual mnemonic 
\be 
\phi _1^{bs} = - {\rm arg} \, V_{cs}^W + {\rm arg} \, V_{cb}^W  
\; , \; 
\phi _2^{bs} = - {\rm arg} \, V_{ub}^W 
\; , \; 
\phi _3^{bs} = 
\pi - {\rm arg} \, V_{cb}^W + {\rm arg} \, V_{ub}^W 
+ {\rm arg} V_{cs}^W 
\ee 
The angle $\phi _1^{bs}$ 
coincides with an angle of the $tu$ triangle 
\be 
\phi _1^{bs} =  \phi _2^{tu} \; , 
\ee
since 
\be 
\frac{V^*_{us}V_{ub}}{V^*_{ts}V_{tb}} = 
\frac{|V_{us}|^2}{|V_{tb}|^2}
\frac{V^*_{tb}V_{ub}}{V^*_{ts}V_{us}}
\ee
and with one of the $bd$ triangle to leading order:
\be 
\phi _2^{bs} = \phi _3 + {\cal O}(\lambda ^2) 
\ee

A more relevant observation is that $\phi _3^{bs}$ can be 
extracted by comparing  
$\bar B_s(t) \to D^+_s K^-$ with $B_s(t) \to D^-_s K^+$ 
\cite{KHOZE,BS4}
with the difference being controlled by 
\be  
{\rm Im} \frac{V^*_{ts}V_{tb}}{V^*_{tb}V_{ts}} 
\frac{V^*_{ub}V_{cs}}{V^*_{us}V_{cb}}\simeq 
\left| \frac{V_{ub}V_{cs}}{V_{us}V_{cb}}
\right| \sin (\phi _2^{bs} + \phi _1^{bs}) = 
\left| \frac{V_{ub}V_{cs}}{V_{us}V_{cb}}
\right| \sin \phi _2^{bs} + {\cal O}(\lambda ^2) 
\ee   
Likewise we find that the difference between 
$\bar B_s(t) \to D_s^-K^+$ and $ B_s(t) \to D_s^+K^-$ is 
given by 
\be 
{\rm Im} \frac{V^*_{ts}V_{tb}}{V^*_{tb}V_{ts}} 
\frac{V^*_{us}V_{cb}}{V^*_{ub}V_{cs}} \simeq 
\left| \frac{V_{us}V_{cb}}{V_{ub}V_{cs}}
\right| \sin (-\phi _2^{bs} - \phi _1^{bs}) = -
\left| \frac{V_{us}V_{cb}}{V_{ub}V_{cs}}
\right| \sin \phi _2^{bs} + {\cal O}(\lambda ^2)  
\ee

\item 
The $tc$ triangle 
\be 
1+ \frac{V^*_{tb}V_{cb}}{V^*_{ts}V_{cs}} + 
\frac{V^*_{td}V_{cd}}{V^*_{ts}V_{cs}} = 0 
\ee 
also has a small angle 
\be 
\phi _1^{tc} = {\rm arg} \left( 
\frac{ V^*_{tb}V_{cb}}{V^*_{ts}V_{cs}} \right)  - \pi = 
{\rm arg} \left( \frac{V^*_{cs}V_{cb}}
{-V^*_{ts}V_{tb}}\right)  = 
\phi _1^{bs} =  {\cal O}(\lambda ^2)
\ee
that coincides with the small 
angle of the $bs$ triangle and 
\be 
\phi _2^{tc} = 
{\rm arg} \left( \frac{V^*_{td}V_{cd}}{-V^*_{tb}V_{cb}}\right) , 
\; 
\phi _3^{tc} = 2\pi -  
{\rm arg} \left( \frac{V^*_{td}V_{cd}}{-V^*_{ts}V_{cs}}\right)  
\simeq  \pi -\phi _2^{tc} 
\ee
\be 
\phi _1^{tc} =  -{\rm arg} \, V_{cs}^W +  
{\rm arg} \, V_{cb}^W  
\; , \; 
\phi _2^{tc} = - {\rm arg} \, V_{cb}^W - 
{\rm arg} \, V_{td}^W 
\; , \; 
\phi _3^{tc} = \pi  +{\rm arg} \, V_{td}^W +  
{\rm arg} \, V_{cs}^W  
\ee 
Yet in the absence of $T_c = [t \bar c]$ mesons this triangle 
does not suggest new ways to measure 
$\phi _1$, $\phi _2$, $\phi _3$ or $\chi$. 
\item 
The $cu$ triangle 
\be 
1 + 
\frac{V^*_{ub}V_{cb}}{V^*_{us}V_{cs}} + 
\frac{V^*_{ud}V_{cd}}{V^*_{us}V_{cs}} = 0 
\ee 
is even more extreme with a tiny angle 
\be 
\phi ^{cu}_3 = 
{\rm arg} \left( 
\frac{-V^*_{ud}V_{cd}}{ V^*_{us}V_{cs}} 
\right) \simeq  - \lambda ^4 A^2 \eta \simeq - 1.6 \cdot 
10^{-3} \eta 
\ee 
Again, it is a novel angle and therefore we denote it 
specifically \cite{PDG2000}:
\be 
\phi _3^{cu} \equiv \chi ^{\prime} 
\mlab{CHIP} 
\ee
Its value can be inferred from \cp~asymmetries in $D$ decays, 
as described in Sect. \ref{CHARM}. 

As in previous cases the other angles practically coincide with 
angles already encountered, namely 
\be 
\phi _1^{cu} =  \pi - 
{\rm arg} \left(\frac{V^*_{ub}V_{cb}}{ V^*_{us}V_{cs}}\right)  
\simeq \pi - \phi _3 + {\cal O}(\lambda ^4) \; , \; 
\phi _2^{cu} = {\rm arg} \left(\frac{-V^*_{ub}V_{cb}}
{ V^*_{ud}V_{cd}}\right) 
\simeq \pi - \phi _1^{cu}  
\simeq \phi _3 
\ee 
\be 
\phi _1^{cu} = \pi - {\rm arg} \, V_{cb}^W 
+ {\rm arg} \, V_{ub}^W +{\rm arg} \, V_{cs}^W, \; 
\phi _2^{cu} = - {\rm arg} \, V_{ub}^W +  
{\rm arg} \, V_{cb}^W , \; 
\phi_3^{cu} = -{\rm arg} \, V_{cs}^W  
\ee 
\item 
The $sd$ triangle 
\be 
1 + \frac{V^*_{ud}V_{us}}{V^*_{cd}V_{cs}} + 
\frac{V^*_{td}V_{ts}}{V^*_{cd}V_{cs}}  
 = 0 
\ee 
has a very similar shape with 
\ba 
\phi _1^{sd} &=& 2\pi - {\rm arg} 
\left( \frac{-V^*_{td}V_{ts}}{ V^*_{cd}V_{cs}}\right) 
\simeq \pi - \phi _2^{sd} \nn
\phi _2^{sd} &=& {\rm arg} 
\left( \frac{V^*_{td}V_{ts}}{-V^*_{ud}V_{us}}\right) \nn 
\phi _3^{sd} &=&  {\rm arg} 
\left( \frac{V^*_{ud}V_{us}}{V^*_{cd}V_{cs}}\right) -\pi = 
\phi _3^{cu}  
\ea 
\be 
\phi _1^{sd} = \pi + {\rm arg} \, V_{td}^W +  
{\rm arg} \, V_{cs}^W 
\; , \; 
\phi _2^{sd} = -{\rm arg} \, V_{td}^W \simeq \phi_1
\; , \; 
\phi _3^{sd} = -{\rm arg} \, V_{cs}^W 
\ee
Obviously it describes \cp~violation in the decays 
of strange hadrons. 
\end{enumerate} 
To summarize this overview of the rescaled unitarity 
triangles: 
\begin{itemize}
\item 
They fall into three groups of two members each:
\begin{enumerate}
\item 
the lengths of all sides are of the same order of magnitude; 
\item 
one is down by ${\cal O}(\lambda ^2)$ relative to the 
other two; 
\item 
one is down by ${\cal O}(\lambda ^4)$. 
\end{enumerate}
\item 
They can be described in terms of four {\em independant} 
angles; two can be picked from the first group, 
\mref{ANGLES1},  and one each 
from the second and third groups, \mref{CHI} and 
\mref{CHIP}, respectively. 
\item 
To {\em leading} order in $\lambda$ both triangles in 
the first group have angles 
$\phi _1$, $\phi _2$ and $\phi _3$; the triangles in 
the second group have angles  
$\chi$, $\phi _1$, $\pi - \phi _1$ and 
$\chi$, $\phi _3$, $\pi - \phi _3$, respectively; in the 
third group $\chi ^{\prime}$, $\phi _1$, $\pi - \phi _1$ and 
$\chi ^{\prime}$, $\phi _3$, $\pi - \phi _3$, respectively. 
The four basic angles referred to above can then be taken 
as $\phi _1$, $\phi _3$, $\chi $ and $\chi ^{\prime}$ 
with $\chi$ and $\chi ^{\prime}$ being small and tiny, 
respectively. 
\item 
All six triangles exhibit a different shape once one goes 
beyond the leading order in $\lambda$. 
\item 
Information contained in the KM matrix is encoded in these 
six  triangles in a highly overconstrained form. It would be 
desirable to determine the angles of all triangles 
with an accuracy of better than ${\cal O}(\lambda ^2)$. 
This, however, is not a realistic goal, also for systematic 
reasons: on top of theoretical uncertainties in 
evaluating hadronic matrix elements -- we will face this 
problem in our later discussion -- we cannot, even in principle, 
study transitions of top {\em hadrons} \cite{TOP}. 
\item 
Nevertheless the experimental information 
that can be inferred for 
sides and triangles is still considerably overconstrained. 
\item 
In the subsequent discussion we will focus on the $bs$ 
triangle in general and on the angles 
$\chi$ and $\chi ^{\prime}$ in particular.

\end{itemize}

\section{The $bs$ Triangle \& $B_s \to \psi  \eta \, , \, 
\psi \phi$}
\label{BS}

The most intriguing angle of the $bs$ triangle -- 
$1 + \frac{V^*_{cs}V_{cb}}{V^*_{ts}V_{tb}} + 
\frac{V^*_{us}V_{ub}}{V^*_{ts}V_{tb}}  
 = 0 $ -- is 
the Cabibbo suppressed quantity 
\be 
\chi \equiv  {\rm arg} (V^*_{cs} V_{cb}/V^*_{ts}V_{tb}) = 
 \lambda ^2 \eta \simeq  \frac{1}{20} \cdot \eta 
\ee
There are several important aspects to this angle: 
\begin{itemize}
\item 
It can be measured through the \cp~asymmetry in 
$B_s(t) \to \psi \eta$ {\em without hadronic pollution} 
\cite{BS1,BS2}: 
\ba 
\Gamma (B_s(t) \to \psi \eta) &\propto&  
e^{- t/\tau _{B_s}} \left( 1 - \sin (\Delta m_s t) \cdot 
{\rm Im} \frac{q}{p} \bar \rho (B_s \to \psi \eta ) \right) \\ 
\Gamma (\bar B_s(t) \to \psi \eta) &\propto&  
e^{- t/\tau _{B_s}} \left( 1 + \sin (\Delta m_s t) \cdot 
{\rm Im} \frac{q}{p} \bar \rho (B_s \to \psi \eta ) \right) \; , 
\ea  
where 
\be  
\bar \rho (B_s \to \psi \eta ) = 
\frac{T(\bar B_s \to \psi \eta ) }{ T(B_s \to \psi \eta ) } 
\ee 
and we find  
\be 
{\rm Im} \frac{q}{p} \bar \rho (B_s \to \psi \eta ) \simeq 
{\rm Im} \frac{\left( V_{tb}V^*_{ts}V_{cs}V^*_{cb}
\right)^2}{\left|   V_{tb}V^*_{ts}V_{cs}V^*_{cb}
\right|^2} = - \sin 2\chi  \simeq -\frac{1}{10} \eta 
\ee 
\item 
This prediction is very reliable in terms of the parameter 
$\eta$. We also know for sure that it is small since 
Cabibbo suppressed by $\lambda ^2$. Numerically one 
estimates at present 
\be 
\sin 2 \chi \simeq (2 \div 5) \%
\ee 
The smallness of this effect is specific to the KM ansatz: in 
$B_s \to \psi \eta$ and $B_s \leadsto \overline B_s \to 
\psi \eta$ the leading contribution involves quarks of the 
second and third families only; yet then no asymmetry can arise 
on this level since the KM ansatz requires the interplay of 
three families (at least). 
\item 
On the other hand 
\cp~violation {\em not} connected to the family structure will 
{\em not} be reduced here. Comparing $B_s(t) \to \psi \eta$ 
and $\overline B_s(t) \to \psi \eta$ thus provides both a 
{\em promising} and a {\em clean} laboratory to search for a 
manifestation of New Physics.  

\end{itemize}
Thus there are two aspects to probing $B_s \to \psi \eta$: 
\begin{itemize} 
\item 
Any asymmetry in this channel that exceeds a few percent 
is a clear manifestation of New Physics. That means if an 
asymmetry is to become observable in the next few years, 
New Physics has to intervene. 
The KM prediction will be made more precise in the 
foreseeable future through data on $B_d \to \psi K_S$ and 
$|V_{ub}/V_{cb}|$.  
\item 
Even if no asymmetry is found above the 
KM expectation, it would 
be important to probe the region below that level as an essential 
self-consistency check on the completeness of the KM ansatz. 
For the three sides of the $bs$ triangle can be determined and 
also (hopefully) $\phi _3^{bs} \simeq \phi _3$ in addition 
to $\chi$. 
\end{itemize}
At this point (if not before) it becomes mandatory to deal 
with a technical complication: while the final state 
$\psi \eta$ is a pure \cp~eigenstate, $\psi \phi$ is not: 
for an S-[or D-]wave configuration it is \cp~even, 
for a P-wave it is odd. While the S-wave is expected to 
dominate due to kinematics, the P-wave will be present as 
well with an asymmetry of equal size, yet {\em opposite} 
sign! To avoid this (partial) compensation, which one might 
ill afford, one has to unfold the S- and P-wave components 
\cite{DUNIETZ}.   

The angle $\chi$ controls also the \cp~asymmetry in 
semileptonic $B_s$ decays: 
\ba  
a_{SL} (B_s) &=&  
\frac{\Gamma (\bar B_s(t) \to l^+ X) - \Gamma ( B_s(t) \to l^- X)} 
{\Gamma (\bar B_s(t) \to l^+ X) + \Gamma ( B_s(t) \to l^- X)} \nn 
&= &
\frac{\left| \frac{q}{p}\right| ^2 - \left| \frac{p}{q}\right| ^2} 
{\left| \frac{q}{p}\right| ^2 + \left| \frac{p}{q}\right| ^2} = 
\frac{\Delta \Gamma (B_s)}{\Delta M(B_s)} 
\sin \Phi (\Delta B=2) \nn   
\sin \Phi (\Delta B=2) &\simeq& F \cdot \frac{m_c^2}{m_b^2} 
\sin \chi 
\ea  
where one estimates 
\be 
F \sim 3 
\mlab{FUDGE} 
\ee 
and thus 
\be 
a_{SL} (B_s) \sim {\cal O}(10^{-4}) 
\ee
This prediction lacks numerical precision -- \mref{FUDGE} 
represents a rough estimate only and $\Delta \Gamma (B_s)/
\Delta M(B_s)$ is not known (yet) -- and there is thus no 
realistic hope to extract sin $\chi$ from it. 

However the small value of $\sin \Phi (\Delta B=2)$ is again very 
specific to the KM ansatz and New Physics could enhance it 
greatly to the 1 \% level. In that case the asymmetry in 
$B_s \to \psi \eta$ would be likewise enhanced. Studying 
$a_{SL}(B_s)$ thus represents a back-up option in case 
$B_s \to \psi \eta$, $\psi \phi$ cannot be analyzed with the 
required sensitivity. 

\section{Unitarity Triangle for Charm Decays}
\label{CHARM} 

The $cu$ triangle -- 
$1 + 
\frac{V^*_{ub}V_{cb}}{V^*_{us}V_{cs}}  
+ \frac{V^*_{ud}V_{cd}}{V^*_{us}V_{cs}} = 0$ -- has a very 
extreme shape: 
\be 
\left| \frac{V^*_{ud}V_{cd}}{V^*_{us}V_{cs}} \right| = 
1+ {\cal O}(\lambda ^4) \; , \; \; 
\left| \frac{V^*_{ub}V_{cb}}{V^*_{us}V_{cs}} \right| 
\sim {\cal O}(\lambda ^4) 
\ee 
and a tiny angle 
\cite{CHIPCHARM} 
\be 
\chi ^{\prime} = {\rm arg} \left( 
\frac{V^*_{ud}V_{cd}}{V^*_{us}V_{cs}} \right) \simeq  
A^2 \lambda ^4 \eta \simeq 1.6 \cdot 10^{-3} \eta 
\mlab{CHIPRIMEV}
\ee 

First we list decay modes 
that would provide us with access to this new angle; then we 
will address the more complex issue of how well one might 
do in dealing with hadronic pollution to extract a numerical 
value or, alternatively, how we can reliably identify the 
intervention of New Physics. 

In describing non-leptonic $D$ decays one can draw 
several different looking diagrams; they are usually referred 
to as spectator, Penguin and weak annihilation processes. 
On top of that one has to allow for prominent final state 
interactions since charm decays proceed in a region populated 
by many resonances; since the final state interactions 
mix those processes it makes little sense to treat them 
individually. Instead we will keep separate contributions 
controlled by different combinations of KM parameters while 
lumping all processes with the same KM dependance into one 
amplitude. 

One has to distinguish between two categories, namely 
\cp~violation involving $D^0 - \bar D^0$ oscillations 
and direct \cp~violation. The former allows for an almost 
zero background search for New Physics since both 
$D^0 - \bar D^0$ oscillations as well as \cp~phases are predicted 
to be quite small within the KM ansatz; their combined 
effect is thus truly tiny. Therefore we will discuss the 
latter where the prospects for observing KM effects are 
significantly better (though not good). 

For direct \cp~violation to become observable in 
integrated rates, two amplitudes with different weak and 
strong phases have to contribute. The former implies 
-- within the KM ansatz -- that one has to consider 
Cabibbo suppressed channels; the latter is typically 
satisfied when two different isospin amplitudes 
contribute.

\subsection{$D \to \pi \pi \, , \; K \bar K $}

With $D^{\pm} \to \pi ^{\pm} \pi ^0$ 
described by a single isospin 
amplitude, a \cp~asymmetry can arise only in 
$D^0 \to \pi ^+ \pi ^-$ [and a compensating one in 
$D^0 \to \pi ^0 \pi ^0$] where the final state can be 
$I=0$ or $2$ or in $D^0 \to K^+ K^-$ with 
$I_f = 0,1$. We have 
\be 
T(D^0 \to \pi ^+\pi ^-) \propto 
V_{cd}V^*_{ud} e^{i\delta _2} |T_2| + 
 e^{i\delta _0}
\left( V_{cd}V^*_{ud} |T_0| + 
V_{cs}V^*_{us} |\tilde T_0| 
\right) \; , 
\ee 
where $T_2$ [$T_0, \, \tilde T_0$] denote the transition 
amplitudes for $I=2,\, 0$ final states with the CKM parameters 
and strong phase shifts $\delta _2$ [$\delta _0$] factored 
out. In a {\em naive} diagramatic representation 
$T_2$ is generated by the spectator process alone and 
$\tilde T_0$ solely by Penguin dynamics whereas $T_0$ 
receives contributions also from Penguin and weak 
annihilation transitions.

For the difference between \cp~conjugate rates one 
then finds  
$$  
\frac{\Gamma (D^0 \to \pi ^+ \pi ^-) - 
\Gamma (\bar D^0 \to \pi ^+ \pi ^-)}
{\Gamma (D^0 \to \pi ^+ \pi ^-) +  
\Gamma (\bar D^0 \to \pi ^+ \pi ^-)} = 
$$ 
\be 
- \frac{2 \sin (\delta _0 - \delta _2) 
{\rm Im} \frac{V_{cs}V^*_{us}}{V_{cd}V^*_{ud}}}
{
1+|r |^2 + |\tilde r |^2 - 2| r \tilde r | 
+2 \cos (\delta _0 - \delta _2) |r | - 
2 \cos (\delta _0 - \delta _2) |\tilde r |
} \; , 
\mlab{DIRECTAS} 
\ee 
with 
\be 
r \equiv \frac{T_0}{T_2} \; \; , \; \; 
\tilde r \equiv \frac{\tilde T_0}{T_2} 
\ee
\mref{DIRECTAS} shows explicitely that one 
needs nontrivial final state interactions 
-- $\delta _0 - \delta _2 \neq 0$ -- and a 
KM phase 
-- Im$\frac{V_{cs}V^*_{us}}{V_{cd}V^*_{ud}} 
\sin \chi ^{\prime} \simeq 
- A^2 \lambda ^4 \eta \leq 10^{-3} 
\neq 0$ -- 
for such an asymmetry to become observable.  
>From \mref{CHIPRIMEV} we infer that one expects an 
asymmetry of order $0.1$ \%. It is not inconceivable that 
such an effect might become observable in the future 
allowing us to probe the value of $\chi ^{\prime}$.

However this number can serve as 
an approximate guideline only since the actual value depends 
on the size of hadronic matrix elements contained in 
$r$ and $\tilde r$ which 
are notoriously hard to evaluate. Thus at present one could 
{\em not} rule out {\em with certainty} that a value as 
"high" as $0.5$ \% or so 
could be accommodated within the KM ansatz.

\subsection{$D^{\pm} \to K_{S,L} \pi ^{\pm}$}

The channels $D^+ \to K_S \pi ^+$ as well as $D^+ \to K_L \pi ^+$ 
can exhibit a \cp~asymmetry since they involve the interference 
between the Cabibbo allowed $D^+ \to \bar K^0 \pi ^+$ and 
doubly Cabibbo suppressed $D^+ \to K^0 \pi ^+$ channels 
\cite{HITOSHI}. 
There are actually two distinct sources for a \cp~asymmetry here, 
namely 
\begin{itemize}
\item 
in the $\Delta C = 1$ sector we are primarily interested in and 
\item 
in $\Delta S=2$ dynamics generating the $K_S$ (and $K_L$) 
mass eigenstate from the $\bar K^0$ and $K^0$ flavour 
eigenstates. 
\end{itemize} 
Ignoring the latter one finds for the transition amplitude 
\be 
T(D^+ \to K_S \pi ^+) = 
V_{cs}V^*_{ud} \left( \hat T_1 + 
\frac{V_{cd}V^*_{us}}{V_{cs}V^*_{ud}} \hat T_2 
\right) 
\ee
Since 
\be 
\frac{V_{cd}V^*_{us}}{V_{cs}V^*_{ud}} = 
\frac{|V_{us}|^2}{|V_{ud}|^2} \cdot 
\frac{V_{cd}V_{ud}}{V_{us}V_{cs}}
\ee 
we have 
\footnote{
At first sight it would seem that the quantities 
$\frac{V_{cd}V^*_{us}}{V_{cs}V^*_{ud}}$ or 
$\frac{V_{cd}V_{ud}}{V_{us}V_{cs}}$ cannot be observables 
since they are not rephasing invariant. However the 
$s$ and $d$ phases can be absorbed into the $K$ 
state vectors -- as it happens in the more familiar 
case of $\frac{V_{tb}V^*_{td}}{V_{td}V^*_{tb}} \cdot 
\frac{V_{cs}V^*_{cb}}{V_{cb}V^*_{cs}}$ 
which controls the \cp~asymmetry in $B_d \to \psi K_S$. } 
\be 
\Im \frac{V_{cd}V^*_{us}}{V_{cs}V^*_{ud}} = 
- \lambda ^2 \chi ^{\prime} \simeq 
- \eta A^2 \lambda ^6 \simeq - 8 \cdot 10^{-5} \eta 
\ee 
It seems unrealistic that such a tiny effect could ever be 
measured. 

Within the KM ansatz the only observable effect is due to 
\cp~violation in the $K^0 - \bar K^0$ complex: 
$$  
\frac{\Gamma (D^+ \to K_S \pi ^+) - \Gamma (D^- \to K_S \pi ^-)}
{\Gamma (D^+ \to K_S \pi ^+) + \Gamma (D^- \to K_S \pi ^-)} 
\simeq - 2 {\rm Re} \epsilon _K \simeq 
- 3.3 \cdot 10^{-3} \simeq 
$$ 
\be 
\simeq \frac{\Gamma (D^+ \to K_L \pi ^+) - 
\Gamma (D^- \to K_L\pi ^-)}
{\Gamma (D^+ \to K_L \pi ^+) + \Gamma (D^- \to K_L\pi ^-)} 
\ee 
The real lesson to be learnt here is the following: 
\begin{itemize}
\item 
Comparing $D^+ \to K_{S,L} \pi ^+$ and 
$D^- \to K_{S,L} \pi ^-$ provides an almost zero background probe 
for New Physics which could very conceivably enter through the 
doubly  Cabibbo suppressed amplitude for $D^+ \to K^0 \pi ^+$ 
\cite{NIR}.  
\item 
The intervention of New Physics can be distinguished against the 
effect driven by $\epsilon _K \neq 0$ through the size of the 
asymmetry and its relative sign in the $K_L \pi ^{\pm}$ and 
$K_S \pi ^{\pm}$ final states. For New Physics generates 
\be 
\frac{\Gamma (D^+ \to K_S\pi ^+) - 
\Gamma (D^- \to K_S\pi ^-)}
{\Gamma (D^+ \to K_S \pi ^+) + \Gamma (D^- \to K_S\pi ^-)} = 
- \frac{\Gamma (D^+ \to K_L \pi ^+) - 
\Gamma (D^- \to K_L\pi ^-)}
{\Gamma (D^+ \to K_L \pi ^+) + \Gamma (D^- \to K_L\pi ^-)} 
\ee 

\end{itemize} 

\section{Gateways for New Physics}
\label{GATES} 

The corrollary to testing the completeness of the KM description 
is to search for manifestations of New Physics. A comprehensive 
program analyzing $B$ decays can also reveal salient features 
of that New Physics in addition to its existence. 

New Physics is most likely to enter 
$\Delta B=2$ and $\Delta C=2$ dynamics driving 
$B^0 - \bar B^0$ and $D^0 - \bar D^0$ oscillations.  
Suppressed $\Delta B=1$ and $\Delta C=1$ decays are promising 
as well, in particular if the transition is dominated by a 
Penguin operator offering access to high mass scale dynamics. 

We will illustrate this briefly through some examples: 
\begin{itemize} 
\item 
New Physics contributing significantly or even dominantly 
to $B_s - \bar B_s$ oscillations could 
\begin{itemize}
\item 
lead to 
\be 
\phi _1 + \phi _2 + \phi _3 \neq 0 \; , 
\ee 
\item 
enhance the \cp~asymmetry in $B_s(t) \to \psi \eta$ and in 
$B_s \to l^- X$ by an order of magnitude even and 
\item 
cause a different value of $\phi _3$ to be extracted 
from $B^{\pm} \to D_{\pm}K^{\pm}$ and 
$B_s(t) \to D_s^+ K^-$.  
\end{itemize}
\item 
If New Physics contributed to $B_d - \bar B_d$ oscillations, 
the \cp~asymmetry measured in $B_d \to \psi K_S$ would 
probably differ from the value of $\phi _1$ inferred from the 
$bd$ triangle constructed through its sides. 
\item 
New Physics in the strong Penguin transition 
$b \to s q \bar q$ could induce a significant difference 
in the values obtained for $\phi _1$ from the 
\cp~asymmetries in $B_d \to \psi K_S$ and 
$B_d \to \phi K_S$. 
\item 
New Physics could induce a {\em time dependant} 
\cp~asymmetry 
\footnote{This asymmetry should not be confused with a 
{\em direct} \cp~asymmetry.} 
in $D^0 \to K^+K^-$, $\pi ^+ \pi ^-$, 
$K_S \pi ^0$, $K_S \eta $ etc. of a few percent even and 
an order of magnitude larger in $D^0 \to K^+ \pi ^-$. 
The KM background is completely negligible. 
\item 
New Physics intervening in doubly Cabibbo suppressed 
channels could induce direct \cp~asymmetry in 
$D^{\pm} \to K_{S,L}\pi ^{\pm}$ of a few percent - 
again with insignificant KM background.   

\end{itemize} 

Some general comments should be made concerning 
$\Delta C \neq 0$ vs. $\Delta B \neq 0$ dynamics: 
\begin{itemize}
\item 
A priori it is quite conceivable that {\em qualitatively} 
different forces drive the decays of down- and up-type 
quarks. More specifically, non-Standard-Model forces 
might exhibit a very different pattern for the two classes 
of quarks. 
\item 
The `background' due to Standard Model forces is in general 
higher in the decays of up-type than down-type quarks since the 
former are KM allowed whereas the latter are KM forbidden 
transitions. 
\item 
Charm decays then offer not only the best, but probably  
a quite unique window onto this landscape: while on 
one hand nonstrange 
light-flavour hadrons do not allow for oscillations, 
doubly Cabibbo suppressed transitions and other rare decay, 
top {\em hadrons} on the other hand do not even form in a 
practical way.
\item 
It is thus conceivable that even detailed studies of beauty 
decays might not reveal the intervention of New Physics, 
yet charm decays will!   

\end{itemize} 

\section{Summary and Outlook}
\label{SUMMARY} 

Some of our findings are of a qualitative and others of a more 
quantitative nature: 
\begin{itemize}
\item 
The six unitarity triangles (and 
the three weak universality relations) represent the information 
contained in the KM matrix in an immensely overconstrained 
form. 
\begin{itemize}
\item 
Fourteen out of the total of eighteen angles are naturally large; 
two are 
of order $\lambda ^2$ and the two remaining ones of order 
$\lambda ^4$. To leading order in $\lambda$ those fourteen 
large angles coincide into just two independant angles and 
their complements. In order $\lambda ^2$ differences 
emerge between them. These findings are already apparent 
in the structure of the KM matrix, see \mref{WOLFKM}.   
\item 
In principle all angles of the six triangles could be measured 
through \cp~asymmetries in top, beauty, charm and 
strange weak decays. 
\item 
In practise that is not possible for a variety of reasons: 
absence of top hadrons, tiny effective branching ratios, 
theoretical uncertainties in the size of hadronic matrix 
elements etc. 
\item 
Yet the measurements that appear feasible will still provide 
us with a highly overconstrained data set that carries a 
high promise for revealing New Physics.

\end{itemize}
\item 
\cp~studies in $B$ decays will at first proceed in two stages 
that have been discussed extensively in the literature: 
\begin{itemize}
\item 
The first task will be to establish the existence of \cp~violation 
in $B$ decays, most likely through observing 
sin$2\phi _1 \neq 0$. 
To increase the size of the available sample one 
will put together different channels driven by the same quark 
level transition, namely $B_d \to \psi K_S$, $\psi K_L$, 
$D \bar D$ etc. 
\item 
The next task will be to extract $\phi _2$ and $\phi _3$ where 
one has to face also the theoretical challenge of having to 
deal with more than one transition operator and its 
hadronic matrix elements. 
\begin{itemize}
\item 
As far as $\phi _2$ is concerned one will presuambly tackle 
this problem by studying several channels driven by the 
same quark level transition, namely $B \to \pi 's$. 
\item 
For $\phi _3$ on the other hand one will analyze different 
quark level transitions. Their \cp~asymmetries depend on 
angles that to {\em leading order} in $\lambda$ coincide with 
$\phi _3$. It will be crucial to see whether the different 
transitions yield consistent values of $\phi _3$. 
\end{itemize}
\end{itemize}
\item 
Yet a comprehensive program has to push tests of the 
completeness of the KM description considerably 
further: 
\begin{itemize} 
\item 
Ultimately the goal has to be to go after the 
${\cal O}(\lambda ^2) \sim 5\%$ differences predicted by the 
KM ansatz between different angles that agree to leading order. 
A significant deviation from those predictions would reveal 
New Physics.  
The most promising case 
\footnote{This does not mean it is a promising case, though.} 
is provided by comparing the \cp~asymmetries in 
$B^+ \to D_{\pm}K^+$, $B_s(t) \to K_S \rho ^0$ and 
$B_s (t) \to D_s^+ K^-$: 
\ba 
B^+ \to D_{\pm}K^+ \; \; \; &\Rightarrow& \; \; \; 
{\rm tg} \phi _3 = \frac{\eta}{\rho} 
\left[ 
1 - \half \lambda ^2 \left( 1 - 2 \rho - 
\frac{2 \eta ^2}{\rho} \right) 
\right] \nn 
B_s(t) \to K_S \rho ^0 \; \; \; &\Rightarrow& \; \; \; 
{\rm tg} \phi _3^{tu} = \frac{\eta}{\rho} 
\left[ 
1 - \half \lambda ^2 \right] \nn 
B_s(t) \to D_s^+ K^- \; \; \; &\Rightarrow& \; \; \; 
{\rm tg} \phi _2^{bs} = {\rm tg} \phi _3 
\ea  
\item 
Rather than search for differences predicted to be small 
between two large 
angles one can undertake to measure novel angles that are  
small in the KM description.   
\item 
Such an angle enters in the $bs$ triangle denoted by 
$\chi \sim {\cal O}(\lambda ^2)$. Its value can be extracted in a 
{\em theoretically clean} way from $B_s \to \psi \eta$ and 
$B_s \to \psi \phi$ where in the latter case one has to 
separate the contributions from even and odd angular 
momentum partial waves. The \cp~asymmetry is reliably 
predicted to be parametrically 
\be 
\sin 2 \chi \simeq 2\lambda ^2 \eta 
\mlab{SIN2CHI}
\ee 
which numerically translates into $2 \div 5$ \% at present. 
\begin{itemize}
\item 
Future data on $B_d \to \psi K_S$ vs. $\bar B_d \to \psi K_S$ 
in particular will make the KM prediction of 
\mref{SIN2CHI} more precise numerically. 
\item 
Observing an asymmetry above the expected value establishes the 
presence of New Physics. 
\item 
Every effort should be made to measure an asymmetry in 
$B_s \to \psi \eta$ even if its size appears to be consistent 
with the KM expectation. For with $\chi$ and 
$|V_{ub}/V_{cb}|$ one can construct the rescaled 
$bs$ triangle and read off $\phi ^{bs}_3$; the latter 
can then be compared  
with the value of $\phi _3$ extracted from other 
transitions. From a significant numerical difference 
between the two one can infer New Physics. 
\end{itemize}
\item 
It has to be stressed that these ${\cal O}(\lambda ^2)$ 
effects could be considerably larger due to New Physics. 
\item 
Another new angle enters in the $cu$ triangle, namely 
$\chi ^{\prime}$, which is very small: 
\be 
\chi ^{\prime} \simeq A^2 \lambda ^4 \eta \sim 10^{-3} 
\ee
It will give rise to \cp~asymmetries in 
$D^0 \to \pi ^+ \pi ^-$, $K^* K$, $\rho \pi$ etc. 
with a characteristic scale of $0.1$ \%. The main theoretical 
uncertainty originates from the relative size of the various 
hadronic matrix elements including their phase shifts. On 
general grounds one actually expects considerable variations 
in the size of the observable \cp~asymmetries in the different  
channels around the 0.1 \% level. At present values of 
0.5 \% -- and possibly even 1 \% -- appear conceivable without 
a clear need of New Physics. Three comments can elucidate the 
situation: 
\begin{itemize}
\item 
A dedicated experimental effort should be made to study charm decays 
with a sensitivity level of 0.1 \% for 
\cp~asymmetries. Finding a signal is a fundamental discovery 
irrespective of its theoretical interpretation. 
\item 
Establishing a signal above the 1 \% level provides strong 
evidence for New Physics. 
\item 
Detailed theoretical engineering -- namely describing a host 
of well-measured $D$ decay channels -- might enable us 
to establish whether an observed asymmetry is still consistent 
with KM expectation or reveals New Physics 
\cite{BUCELLA}. 
\end{itemize} 
\item 
The related quantity 
\be 
{\rm Im} \frac{V_{cd}V^*_{us}}{V_{cs}V^*_{ud}} = 
- \lambda ^2 \chi ^{\prime} \simeq 
- \eta A^2 \lambda ^6 \simeq - {\rm few} \times 
10^{-5} 
\ee 
can in principle be probed in $D^{\pm} \to K_{S,L}\pi ^{\pm}$ 
decays. In practise it is so small that there is no realistic 
hope to ever extract it. On the other hand that means that any 
\cp~asymmetry observed here over the well-known and 
precisely predicted one due to $\epsilon _K$ is safely  
ascribed to New Physics.  
\end{itemize} 

\begin{figure}[t]
\epsfxsize=10cm
\centerline{\epsfbox{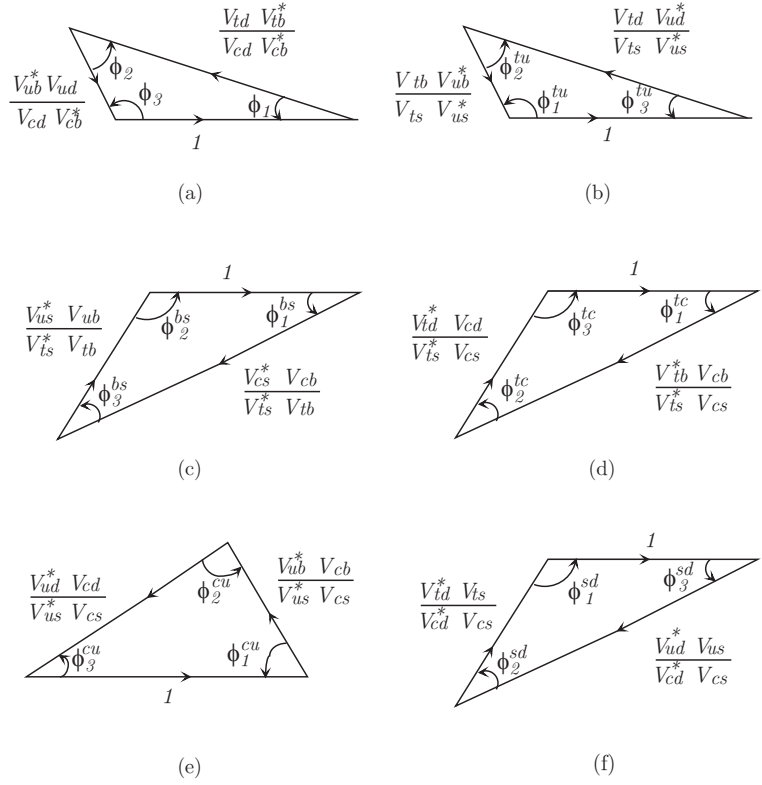}}
\caption{The unitarity triangle for B decays.}
\mlabf{unit2}
\end{figure}

\end{itemize}
To state it in a nutshell: before an experimental program on 
$B^0 - \bar B^0$ oscillations and \cp~violation can be called 
complete it must meet at least the following 
benchmarks: 
\begin{itemize}
\item 
$\phi _1$, $\phi_2$ and $\phi_3$ have been extracted with a relative 
accuracy of better than 5\%. 
\item 
The values of $\phi _3$ inferred from different $B$ transitions 
have been compared and it has been  
analyzed whether the $\lambda ^2$ corrections predicted by 
the KM scheme can be identified. While it is not clear whether 
this is a realistic goal, it has to be attempted nevertheless. 
\item 
It is absolutely mandatory to measure $\chi$ in 
$B_s \to \psi \eta$, $\psi \phi$ with a sensitivity on the 
percent level which carries a high potential to reveal even a 
subtle intervention of New Physics. 
\item 
The angle $\chi ^{\prime}$ induces direct \cp~asymmetries 
in Cabibbo suppressed nonleptonic $D$ decays on about 
the 0.1 \% level. 
Every effort should be made to acquire the experimental 
sensitivity to observe such effects. 
Establishing whether an observed effect indeed is 
consistent with KM or requires New Physics will be a highly 
challenging task unless the signal is an order of magnitude 
larger. Yet it represents an important benchmark 
nevertheless. 
\item 
Finally one has to search for an asymmetry in 
$D^{\pm} \to K_{S,L}\pi ^{\pm}$ decays. Irrespective of the 
origin of \cp~violation an asymmetry has to arise there 
as described by the observable $\epsilon _K$; KM 
$\Delta C=1$ dynamics 
creates an asymmetry described by 
$\lambda ^2 \chi ^{\prime} \leq 10^{-4}$ which in all 
likelihood is too small to be ever observed. Such studies 
thus represent zero background searches for 
New Physics -- analogous to probing for \cp~asymmetries 
in the decay time evolutions in 
$D^0 (t) \to \pi ^+ \pi ^-$, $K^+ K^-$, $K^+ \pi ^-$ transitions 
which involve $D^0 - \bar D^0$ oscillations. 
\item 
The discovery potential for New Physics in heavy flavour 
Dynamics has {\em not} been exhausted, unless a comprehensive 
and dedicated program in charm decays has been pursued! 

\end{itemize}


\vskip 3mm  
{\bf Acknowledgements} 
\vskip 3mm  
The work of I.I.B. has been supported by the NSF under the grant 
PHY 96-0508 and that of A.I.S. by Grant-in-Aid for Special Project 
Research (Physics of CP violation).


\end{document}